\documentclass[10pt]{iopart}

\usepackage{graphicx}
\usepackage[colorlinks=true, allcolors=blue]{hyperref}
\usepackage{mathrsfs}%
\usepackage[title]{appendix}%
\usepackage{xcolor}%
\usepackage{textcomp}%
\usepackage{manyfoot}%
\usepackage{booktabs}%
\usepackage{algorithm}%
\usepackage{algorithmicx}%
\usepackage{algpseudocode}%
\usepackage{listings}%
\usepackage{float}
\usepackage{siunitx}

\newcommand{\electron}{{\textup{e}}}

\begin{document}

\title{Probing Flying-Focus Wakefields}

\author{Aaron~Liberman$^{1,\dagger,*}$, Anton~Golovanov$^{1,\dagger}$, Sheroy~Tata$^{1}$, Anda-Maria~Talposi$^{1}$, and Victor~Malka$^{1,\ddagger}$} 

\begin{indented}
\item[1 ] Department of Physics of Complex Systems, Weizmann Institute of Science, Rehovot 7610001, Israel
\item[$\dagger$ ] These authors contributed equally to this work
\item[* ] aaronrafael.liberman@weizmann.ac.il
\item[$\ddagger$ ] victor.malka@weizmann.ac.il
\end{indented}

\begin{abstract}
Flying-focus wakefields, which can propagate with a tunable velocity along the optical axis, are promising solutions to electron dephasing in laser-wakefield accelerators. This is accomplished by a combination of spatio-temporal couplings and focusing with an axiparabola, a specialized optical element which produces a quasi-Bessel beam. If implemented, dephasingless acceleration would allow for a hitherto unachievable mixture of high acceleration gradients and long acceleration lengths. Here, we conduct an in-depth study of the structure and behavior of such a flying-focus wakefield, through a combination of direct imaging and simulations. We show the stability of the wakefield structures, explore how the wakefield evolves with changes of density, study the effects of ionization on the wakefield structure with a variety of gases, and analyze the importance of the focusing position. These insights shed light onto this novel wakefield regime and bring understanding that is important to the realization of dephasingless acceleration. 
\end{abstract}

\ioptwocol
\frenchspacing

\section{Introduction}
Laser-wakefield accelerators (LWFAs) \cite{Tajima_PRL_1979} are the subjects of considerable interest due to their ability to accelerate electrons with gradients several orders of magnitude stronger than those achievable in conventional accelerators \cite{Joshi_Nature_1984,Modena_Nature_1995}. Over the past two decades, LWFAs have seen rapid development, from the demonstration of quasi mono-energetic electron bunches \cite{Faure_Nature_2004,Geddes_Nature_2004,Mangles_Nature_2004} to the current record of 10 GeV electrons \cite{Picksley_PRL_2024,Rockafellow_PoP_2025}. As LWFAs improve, applications open in fundamental physics research \cite{Mirzaie_NaturePhotonics_2024,LUXE_EurPhysJ_2024}, in free-electron lasers \cite{Wang_Nature_2021,Labat_NaturePhotonics_2022,LaBerge_NaturePhotonics_2024} and in improved radiation therapy \cite{Glinec_Med_Phys_2006,Guo_NatureCommunications_2025}.

Further growth of LWFAs in many of these applications requires the achievement of higher electron energies. This will, for example, enable the probing of strong-field quantum electrodynamic (QED) phenomena with an unprecedented quantum nonlinearity parameter, $\chi$ \cite{LUXE_EurPhysJ_2024}. However, the current scaling laws for LWFAs show significant challenges for this continued growth \cite{Esarey_ReviewOfModernPhysics_2009}. The most significant obstacle to the efficient increase of achievable electron energy in LWFAs is electron dephasing -- the fact that the trapped electrons outpace the wakefield itself, decreasing the accelerating gradient and ultimately ending the acceleration \cite{Joshi_Nature_1984,Esarey_ReviewOfModernPhysics_2009}.  
Proposed solutions for electron dephasing include rephasing, multistage LWFAs, and decreasing the plasma density. Rephasing relies on a shrinking of the wakefield that occurs when the plasma density increases along the laser propagation axis, thus moving the trapped electrons from the decelerating to the accelerating phase \cite{Sprangle_PRE_2001,Guillaume_PRL_2015,Gustafsson_ScientificReports_2024}. In a multistage LWFA \cite{Leemans_PhysicsToday_2009,Steinke_Nature_2016} the acceleration is split up between several separate LWFA stages. The stages can be tuned to a length that is short enough to prevent the electrons from fully dephasing. In the next stage, the electrons again begin in the accelerating phase of the wakefield, thus allowing them to be continuously accelerated. 

Reducing the plasma density, meanwhile, decreases the mismatch between the group velocity of light in the plasma and the velocity of the relativistic electrons \cite{Picksley_PRL_2024,Leemans_NaturePhysics_2006,Gonsalves_PRL_2019,Miao_PRX_2022}. This solution has been responsible for the acceleration of the highest electron energies achieved in LWFA to date \cite{Picksley_PRL_2024,Rockafellow_PoP_2025}. However, decreasing the plasma density also decreases the acceleration gradient of the electrons \cite{Lu_PhysRevSTAB_2007}. This makes the scaling laws for such an approach increasingly unfavorable as the electron energy grows \cite{Lu_PhysRevSTAB_2007}. In addition, to maintain the accelerator structure in the plasma, the beam must remain focused. Thus, it requires guiding the laser pulse through a plasma waveguide in order to avoid beam diffraction \cite{Joshi_Nature_1984,Leemans_NaturePhysics_2006,Picksley_PRL_2024,Gonsalves_PRL_2019}. As the accelerator length grows, this is an increasingly challenging problem. 

To prevent dephasing while maintaining favorable scaling with electron energy and keeping the relative simplicity of a single stage LWFA, it is necessary to modify the propagation velocity of the wakefield, allowing the wakefield to remain phase-locked to the trapped electrons \cite{Caizergues_NaturePhotonics_2020,Palastro_PRL_2020}. Tuning the wakefield velocity requires a change in the velocity of the laser driver, something that can be achieved through the advanced focusing scheme known as flying focus \cite{Caizergues_NaturePhotonics_2020,Palastro_PRL_2020,Froula_NaturePhotonics_2018,Debus_PRX_2019,Sainte-Marie_Optica_2017}. Possible implementations of the flying focus include colliding two tilted pulses \cite{Debus_PRX_2019} and the ``chromatic flying focus''---a combination of angular chromatism and group-delay dispersion \cite{Sainte-Marie_Optica_2017,Froula_NaturePhotonics_2018}. However, for high-power, ultrashort laser-pulses, the most suitable implementation of the flying focus involves the combination of spatio-temporal couplings and an axiparabola \cite{Caizergues_NaturePhotonics_2020}, a reflective long-focal-depth optical element that focuses the laser into a quasi-Bessel beam \cite{Caizergues_NaturePhotonics_2020,Palastro_PRL_2020}. Theoretical works have shown that this combination could accelerate electrons with energies of over 100\,GeV \cite{Caizergues_NaturePhotonics_2020} in operational laser facilities such as ELI-NP \cite{Radier_HPLSE_2022} and in facilities currently under development, such as NSF OPAL \cite{Shaw_Arxiv_2025}. 

Much effort has been devoted to studying this implementation of the flying focus and its feasibility for dephasingless LWFA \cite{Caizergues_NaturePhotonics_2020,Palastro_PRL_2020,Liberman_OL_2024,Ambat_OpticsExpress_2023,Pigeon_OE_2024,Geng_PoP_2022,liberman2025directobservationwakefieldgenerated,Geng_ChinesePhys_2023,Abedi-Varaki_APA_2025,Ramsey_PRA_2023}. A number of numerical studies have been conducted using particle-in-cell (PIC) codes to simulate the wakefield \cite{Caizergues_NaturePhotonics_2020,Palastro_PRL_2020,Geng_ChinesePhys_2023,Abedi-Varaki_APA_2025,Miller_ScientificReports_2023}. In addition to that, experiments have been conducted to demonstrate that the combination of the axiparabola and spatio-temporal couplings gives the ability to manipulate the on-axis propagation of the intensity peak, tuning it to be superluminal, luminal, or subluminal \cite{Liberman_OL_2024,Pigeon_OE_2024}. Recent experimental work has also yielded the first direct observation of the structure of wakefields formed by such pulses \cite{liberman2025directobservationwakefieldgenerated} and has demonstrated that these wakefields are able to maintain the coherent structures necessary for accelerating relativistic electrons \cite{Liberman_SPIE_2025,liberman2025electronaccelerationtunablevelocitylaser}. 

\begin{figure*} [t!]
   \begin{center}
    \includegraphics[width=\linewidth]{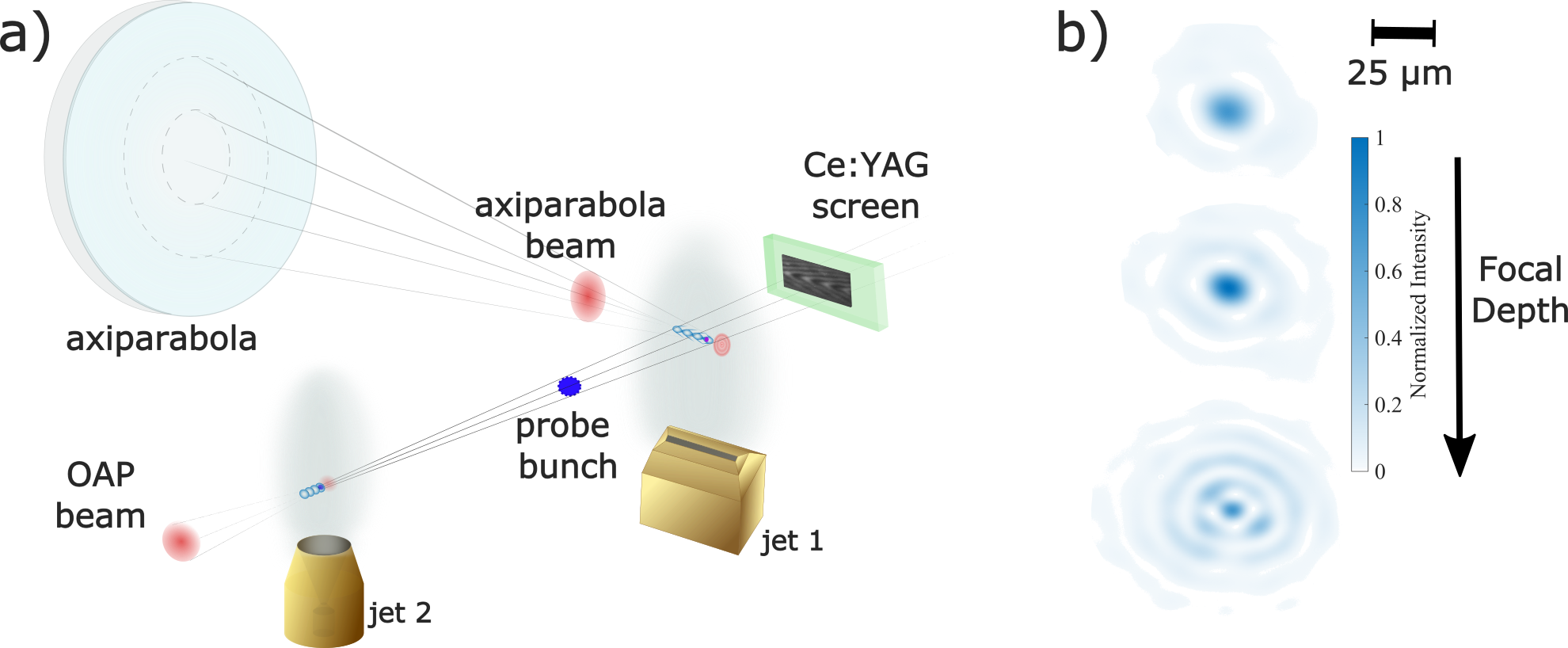}
    \end{center}
    \caption[]{ \label{fig:setup} \textbf{(a) Schematic of the experimental setup. The axiparabola focuses one beam onto jet 1, a slit nozzle, generating a flying-focus wakefield. A second beam, temporally synchronized with the first, is focused by an off-axis parabola (OAP, not shown) onto jet 2, a converging-diverging nozzle, generating a second wakefield. This wakefield accelerates electrons, referred to on the figure as the probe bunch. This bunch is allowed to propagate further, spatially expanding. It then impinges upon the flying-focus wakefield. The fields inside of the wakefield give momentum kicks to the probe electrons. After further free-space propagation, these momentum kicks become density perturbations which can be seen on a Ce:YAG scintillating screen. (b) Three 2D focal spot measurements at 1 mm intervals along the focal line of the axiparabola. Color bar shows relative intensity. Taken in vacuum.}}
\end{figure*} 

We present new insights gained from direct imaging of this flying focus wakefield and PIC simulations. The direct imaging was conducted using femtosecond relativistic electron microscopy (FREM) \cite{Zhang_ScientificReports_2016,Wan_NaturePhysics_2022,Wan_2023_LSA_12_116}, a diagnostic which can probe the electromagnetic field inside of the wakefield with micrometer-scale spatial resolution and femtosecond temporal resolution. The direct imaging serves as a guide for our simulations, reinforcing that we are accurately capturing the essential physics of the wakefield. We experimentally demonstrate the stability of the wakefield structure. Through simulations, we show how the structure of the wakefield changes with different densities, which is a key parameter since the optimal density for this structured wakefield differs significantly for different laser systems \cite{Caizergues_NaturePhotonics_2020}. We examine the influence of different gas mixtures and the importance of fully considering the effects of ionization in the simulation. While these effects may sometimes be neglected in standard, parabola-focused wakefields without introducing significant error, we show that with this flying focus wakefield, ionization effects can significantly alter its structure and behavior. Thus, to correctly simulate the dynamics, full consideration of ionization effects is critical. Finally, we show how the focusing depth inside of the nozzle impacts the wakefield. The insights presented in this work help lay the foundations for the full implementation of the flying-focus, dephasingless LWFA.

\section{Methods}

\subsection{Laser}

The experiment was conducted using the HIGGINS laser system \cite{Kroupp_MRE_2022} at the Weizmann Institute of Science. The HIGGINS system is a Ti:Sapphire laser system that is based on chirped pulse amplification \cite{Strickland_OpticsCom_1985}. The laser was set to provide two 27 fs laser pulses with a central wavelength of 800 nm, an on-target energy of 1 J, and an unfocused diameter for 50 mm. The two laser pulses were split from a common seed pulse. This pulse was stretched and amplified and then split by a 50/50 beam splitter. Each pulse was then individually compressed. This design allowed for the temporal synchronization of the two pulses up to a 10 femtosecond-scale jitter and controllable delay from the different path lengths of the two beams. Each beam was impinged on a deformable mirror. A closed-loop focal spot optimization was performed for each beam. 

\subsection{Flying-Focus Wakefield}

The beam used to generate the flying-focus, shown in figure \ref{fig:setup} (a) as ``axiparabola beam'', was focused by an off-axis axiparabola with an off-axis angle of 10 degrees. The axiparabola had a nominal focal length, $f_0$, of 480 mm (f/9.6) and a focal depth, $\delta$, of 5 mm. The axiparabola introduced a controlled amount of spherical aberrations to the beam, resulting in a distribution of focus as a function of radius of the beam with the form: $f(r) = f_0 + \delta (r/R)^2$, where $r$ is the radial coordinate and $R$ is the total radius of the beam. The axiparabola beam was focused onto a 15 mm long, 4 mm wide supersonic slit nozzle, generating the flying focus wakefield. The nozzle is shown in figure \ref{fig:setup} (a) as “jet 1”. The gas used in the experiment was a 97\% helium, 3\% nitrogen mixture which resulted in a plasma density of around \SI{5e17}{cm^{-3}}, that allowed FREM probing while preventing electron acceleration that could distort the structure of the wakefield. Figure \ref{fig:setup} (b) shows three 2D images of the focal spot of the axiparabola, demonstrating how the focus evolves over the focal depth. 

\subsection{Probe Electron Bunch}

The probe electron bunch, shown in figure \ref{fig:setup} (a) as ``probe bunch'', was generated by focusing a second beam, shown in figure \ref{fig:setup} (a) as ``OAP beam''. The beam was focused by a 1.5 m focal length (f/30) off-axis parabolic mirror which had an off-axis angle of 11 degrees. The beam was focused onto a supersonic converging-diverging nozzle, shown in figure \ref{fig:setup} (a) as ``jet 2'', which had a throat diameter of \SI{500}{\um} and an outlet diameter of 3 mm. The same gas mixture was used, facilitating ionization injection. The measured electron energy was between 100\,MeV and 350\,MeV with a quasi-monoenergetic peak at 270\,MeV $\pm$ 35\,MeV. The measured FWHM divergence was 1.6 mrad. 

\subsection{FREM Diagnostic}

The FREM diagnostic requires precise temporal and spatial alignment of the probe electron bunch with the wakefield under study. The jets were arranged such that the electrons from jet 2 propagated in free space for 10 cm, allowing them to spatially spread out to several hundred micrometers prior to hitting the flying focus wakefield. This was done to allow them to probe a significant section of the plasma in each shot. A motorized mirror allowed for the steering of the OAP beam and thus of the probe electron beam, helping ensure spatial overlap. After passing through the wakefield in jet 1, the probe electron bunch was allowed to drift for 7 mm, thus turning the momentum kicks imparted by the wakefield into local transverse density perturbations in the probe electron bunch. These electrons then impinged on a 30 micrometer thick Ce:YAG screen, allowing the density to be imaged. The screen is shown in figure \ref{fig:setup} (a). 

The initial alignment was done by using the OAP beam as a shadowgraphy probe of the ionization front of the axiparabola beam. Catching the point at which the ionization begins ensured a temporal overlap in the hundreds of femtoseconds and sub-millimeter spatial overlap. Fine alignment was then done with the electron bunch, imaged by the Ce:YAG screen. By adjusting a delay line earlier in the beam line, the temporal delay between the two beams was tuned until the probe electrons illuminated the beginning of the LWFA. 

The light from the Ce:YAG screen was collected by a Mitutoyo plan-apochromatic, infinity-corrected, long-working-distance 10X microscope objective and an achromatic lens with a focal length of 300 mm from Thorlabs. The camera used was a Hamamatsu ORCA-FLASH4.0 digital CMOS camera. Together, the imaging system provided a resolution of \SI{2.2}{\um}, as determined by a 1951 USAF resolution target. The Ce:YAG screen was shielded from infrared noise from the laser with a \SI{100}{\um} thick stainless steel plate.

For the experimental FREM images shown, the spatial dimensions were scaled down by a factor of 1.07 in order to correct for the transverse expansion of the probe after passing through the wakefield. In addition, the experimental images shown were post-processed to remove noise and background noise irradiation.  

\subsection{Beam Propagation Simulations}

To simulate the propagation of the reflected laser pulse from the axiparabola surface to the laser--gas interaction area located in the focus, we used the Axiprop code which solves the wave equation in vacuum \cite{Andriyash_Axiprop,Oubrerie_JoO_2022}.
In order to reduce the computational cost, we used an axisymmetric model of an on-axis axiparabola with the parameters corresponding to the experiment ($f_0 = \SI{480}{\mm}$ focal length, $\delta = \SI{5}{\mm}$ focal depth for $R = \SI{25}{\mm}$ radius). 
Before the reflection, the laser pulse with a central wavelength of \SI{800}{\nm} had a Gaussian temporal profile with a duration of \SI{27}{\fs} (FWHM intensity) and a 6th order super-Gaussian transverse profile with a diameter of \SI{40}{\mm} (FWHM intensity).
The energy of the pulse was equal to \SI{0.85}{\J}.
The lower energy and diameter used in the simulation corresponded better to the experimental observations, perhaps due to non-ideal alignment of the axiparabola and phase-front defects of the experimental pulse.
The simulated field distributions in the focal area were saved in the LASY format \cite{thevenet_lasy} for further use in PIC simulations.

\begin{figure*}[tb!]
   \begin{center}    \includegraphics[width=\linewidth]{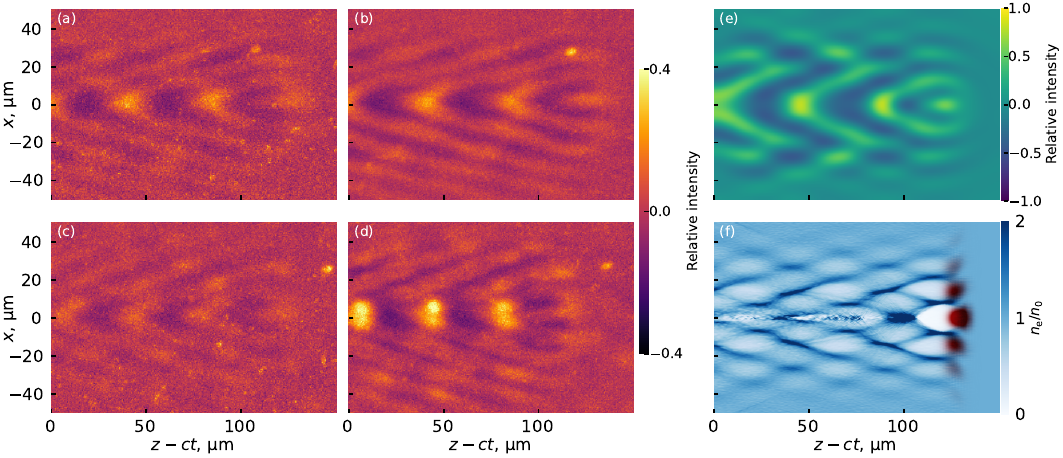}
    \end{center}
    \caption[]{ \label{fig:experiment} \textbf{(a--d) Experimental FREM images. (e) Simulated FREM image with parameters reported in section \ref{sec:methods:PIC} at the plasma density of $n_0 = \SI{5e17}{cm^{-3}}$. (f) Simulated wakefield corresponding to the FREM image. Colorbar in (a--e) shows relative intensity of signal from scintillating screen, where 0 is the intensity of the unperturbed probe beam. Blue colorbar in (f) shows relative electron density distribution $n_\electron/n_0$ and red color shows the intensity of the axiparabola laser field.}}
\end{figure*} 

\subsection{Particle-in-cell Simulation}
\label{sec:methods:PIC}

PIC simulations of laser--plasma interaction were performed with the quasi-3D spectral code FBPIC with azimuthal mode decomposition \cite{Lehe_ComPhysCom_2016}.
The laser pulse was initialized from the field distribution imported from Axiprop simulations assuming linear polarization in the horizontal direction. 
The transverse size of the simulation box was chosen automatically to make sure the entire imported profile fits into the box.
The size in the longitudinal $z$ direction was equal to \SI{150}{\um}, and two azimuthal modes were used.
The grid resolution was $dr = \SI{0.24}{\um}$ and $dz = \SI{0.04}{\um}$, respectively.
To accelerate the computation, simulations were performed in a Lorentz-boosted frame with a Lorentz factor $\gamma = 3$ \cite{Lehe_2016_PRE_94_53305}.

The slit nozzle gas jet was represented by a 10-mm-long longitudinal profile with a 8\,mm plateau and 1\,mm linear up- and downramps.
In various simulations, hydrogen, helium, and helium--nitrogen mixture gases were used.
The gas density corresponded to the stated plasma density, assuming full ionization of hydrogen and helium and ionization of nitrogen up to the 5\textsuperscript{th} level.
Except the case when preionized gas was considered, the gas was initially neutral to properly account for the diffraction of the weaker laser field at large radii where the gas might be only partially ionized.
The gas was initialized with 32 atoms of each element per a 2D cell (2 in $r$, 2 in $z$, 8 in $\theta$ directions, respectively).

\subsection*{FREM Simulation}

To recreate the FREM images, the probe electron bunch's interaction with the electromagnetic field distribution in the wake obtained from PIC simulations was simulated.
The simulated bunch had parameters close to the experimental: an energy spectrum with a uniform energy distribution between \SI{100}{\MeV} and \SI{300}{\MeV}, a duration of \SI{10}{\fs} with a top-hat current profile, an initial transverse source size of \SI{5}{\um}, and a divergence at maximum energy of \SI{1}{mrad}.
The bunch was modeled by \num{5e6} particles of equal weights.
The electromagnetic field was assumed to be moving perpendicular to the beam at the speed of light.

Before interacting with the wakefield, the bunch propagated for \SI{10}{\cm} in vacuum.
After the interaction, it was projected to a plane \SI{7}{\mm} away from the wakefield and normal to the bunch propagation direction.
The projection used a pixel size of \SI{0.52}{\um} (equal to the experimental one).
An additional Gaussian filter with a kernel size of 8 pixels (standard deviation) was applied to the image to model the resolution constraints in the experimental imaging system as well as to reduce the excessive noise introduced by a limited number of particles in the simulations.
To eliminate the effect of varying background brightness, the signal on the plane was calculated relative to the signal in the absence of the wakefield.
Similar to the experimental FREM images, the spatial dimensions on the image plane were scaled down by a factor of 1.07 to account for the beam expansion during its propagation to the screen.

\begin{figure*} [tb!]
   \begin{center}
    \includegraphics[width=\linewidth]{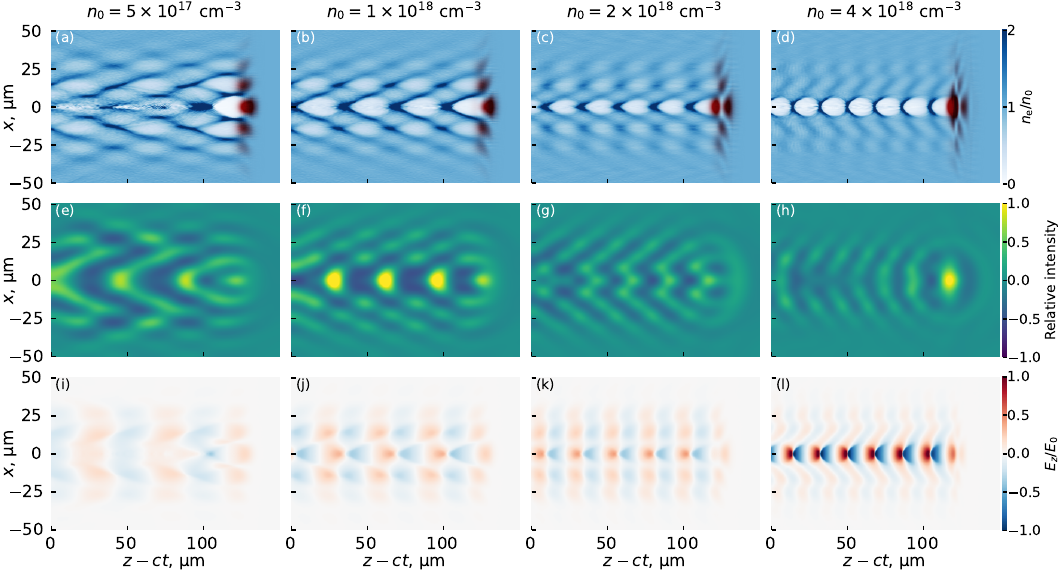}
    \end{center}
    \caption[]{ \label{fig:Density} \textbf{(a--d) Relative electron density distribution $n_\electron/n_0$ and intensity of the axiparabola laser field for flying focus wakefield at different densities. The dashed lines show the FWHM duration of the laser beam as a function of the transverse coordinate. (e--h) Corresponding simulated FREM images at the different densities. Colorbar shows relative intensity of signal from scintillating screen, where 0 is the intensity of the unperturbed probe beam. (i--l) Corresponding spatial distributions of the normalized longitudinal electric field, $E_z/E_0$. Colorbar shows strength of the electric field.}}
\end{figure*}

\section{Results}

\subsection{Measurements of the Flying-Focus Wakefield}

The structure of the flying-focus wakefield is noticeably different from a standard wakefield driven by a parabola-focused pulse. Earlier work \cite{liberman2025directobservationwakefieldgenerated} showed that the structure evolves significantly over the focal depth of the axiparabola and has unique features, such as the presence of significant, radially offset wakefields and the simultaneous mixing of linear and non-linear wakefields \cite{liberman2025directobservationwakefieldgenerated}. 
 
Figure \ref{fig:experiment} (a--d) show four experimental FREM images, taken at the same point in the focal depth and with the same conditions. The images all demonstrate the prominent V-shape structure, shown first in \cite{liberman2025directobservationwakefieldgenerated}. While fluctuations of the probe beam cause changes in the illumination of the wakefield, the structures seen in the wakefield are quite stable over the several shots. The V-structure seen in the FREM images is a combination of the presence of the off-axis wakefields and the propagation of the pulse in plasma, which imposes tilts on the peak intensity profile \cite{liberman2025directobservationwakefieldgenerated}.

Figure \ref{fig:experiment} (e) shows a simulated FREM image that recreates the experimental result. The simulations were performed with a helium gas jet with the plasma density of \SI{5e17}{cm^{-3}} which was inferred from the wakefield period in the experimental FREM images.
In the simulation, the axiparabola focal line began at the depth of $\Delta f_0= \SI{2}{mm}$ after the beginning of the density plateau, and the image in figure \ref{fig:experiment} (e) corresponds to the propagation depth of \SI{4}{mm} after the start of the plateau.
Figure \ref{fig:experiment} (f) shows the simulated wakefield that corresponds to the simulated FREM image.

The close correspondence between the experimental FREM images in figure \ref{fig:experiment} (a--d) and the simulated one in figure \ref{fig:experiment} (e) shows that the simulation is effectively capturing the essential physics involved in this Bessel-like laser beams wakefield. Furthermore, the experimental reproducibility of the structures seen in the FREM images attest to the fact that this is indeed a true representation of the wakefield characteristics, not an accidental occurrence in a single shot. 

\begin{figure*} [t!]
   \begin{center}
    \includegraphics[width=\linewidth]{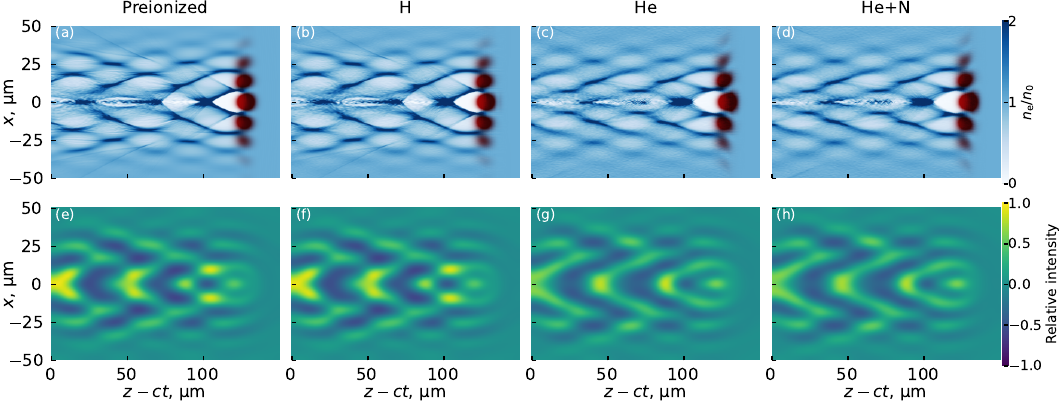}
    \end{center}
    \caption[]{ \label{fig:Ionization} \textbf{Comparison of relative electron density distribution $n_\electron/n_0$ and intensity of the axiparabola laser field for flying focus wakefield using (a) preionized plasma, (b) pure hydrogen, (c) pure helium, and (d) helium-nitrogen mixture. (e--h) Corresponding simulated FREM images. Colorbar shows relative intensity of signal from scintillating screen, where 0 is the intensity of the unperturbed probe beam. }}
\end{figure*} 

\subsection{Structural Change with Density}

A critical parameter in wakefield acceleration in general and in dephasingless LWFA in particular is the plasma density. As was shown in Ref. \cite{Caizergues_NaturePhotonics_2020}, in the flying-focus LWFA scheme, the standard resonance condition $c\tau \approx w_0 \approx \lambda_p$ \cite{Lu_PhysRevSTAB_2007} where $\tau$ is the pulse duration, $w_0$ is the beam waist, and $\lambda_p$ is the plasma wavelength no longer applies. Instead, the resonance condition simplifies to $c\tau \approx \lambda_p$ \cite{Caizergues_NaturePhotonics_2020}. This means that the ideal density for dephasingless LWFA is determined by the pulse duration, and, therefore,  spans a considerable range, between below \SI{e18}{cm^{-3}} to above \SI{5e19}{cm^{-3}}, depending on the laser system. Thus, an exploration of the effect of density on the wake structure generated by Bessel-like beams is critical to the optimization of dephasingless LWFA. 

Figure \ref{fig:Density} (a--d) show simulated wakefields generated with the same simulation parameters as in figure \ref{fig:experiment} (f) but at the densities (a) $n_0=\SI{5e17}{cm^{-3}}$, (b) $n_0=\SI{1e18}{cm^{-3}}$, (c) $n_0=\SI{2e18}{cm^{-3}}$, and (d) $n_0=\SI{4e18}{cm^{-3}}$. The corresponding plasma wavelengths are \SI{47}{\um}, \SI{33}{\um}, \SI{24}{\um}, \SI{17}{\um}, respectively. 

Figure \ref{fig:Density} (e--h) show the corresponding simulated FREM images and figure \ref{fig:Density} (i--l) show the corresponding longitudinal electric fields, $E_z$, normalized by the the cold nonrelativistic wave breaking field $E_0 = \sqrt{m c^2 n_0 / \varepsilon_0}$, where $c$ is the speed of light, $m$ is the electron mass, and $\varepsilon_0$ is the vacuum permittivity. Thus, the plotted field, $E_z/E_0$, accounts for the effect of the plasma density on the expected field. The values of the normalized field, averaged over the accelerating phase, are $0.16$, $0.20$, $0.22$, $0.52$, corresponding to real field values of \SI{11}{GV/m}, \SI{19}{GV/m}, \SI{30}{GV/m}, and \SI{99}{GV/m}, respectively. The increasing values of the normalized field mean that as the density increase, the field becomes more non-linear, with a significant jump for the density of \SI{4e18}{cm^{-3}}. 
In addition, there is the expected shortening of the wakefield period decreases as the density increases. This is due to the fact that the wakefield period is equal to the plasma wavelength which is inversely proportional to $\sqrt{n_0}$ \cite{Chen_Plenum_1983}, and the relativistic wavelength lengthening does not occur for the considered parameters. The impact on the FREM images is a change of the angle of the V-shape pattern due to contraction in the longitudinal direction. As described in Ref. \cite{liberman2025directobservationwakefieldgenerated}, this makes the FREM diagnostic a good measurement of the plasma density \textit{in situ}.

\begin{figure*} [t!]
   \begin{center}
    \includegraphics[width=\linewidth]{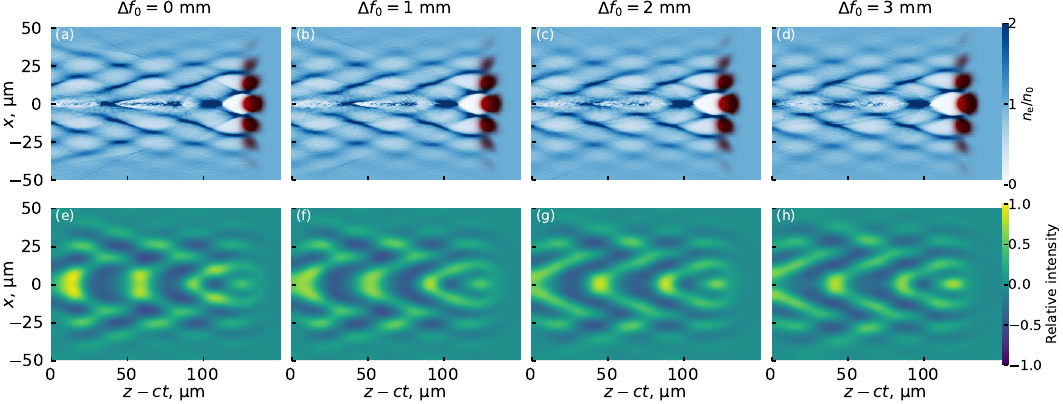}
    \end{center}
    \caption[]{ \label{fig:focusing} \textbf{(a--d) Comparison of relative electron density distribution $n_\electron/n_0$ and intensity of the axiparabola laser field for flying focus wakefield at different focusing locations relative to  beginning of the plateau in the gas density distribution. The simulation iterations are adjusted relative to the change in $\Delta f_0$ and correspond to the position of 2\,mm after the beginning of the focal line. (e--h) Corresponding simulated FREM images. Colorbar shows relative intensity of signal from scintillating screen, where 0 is the intensity of the unperturbed probe beam.}}
\end{figure*} 

The propagation in denser plasma also has a profound effect on the laser driver. As seen in figure \ref{fig:Density} (c) and even more so in (d), propagation through the denser plasma causes the laser driver to split into two distinct pulses on axis. Off axis, the density has a significant impact on the peak intensity profile as well. As was reported in Ref. \cite{liberman2025directobservationwakefieldgenerated}, propagation of the flying focus pulse through the plasma imposes tilts on the peak intensity profile of laser. These tilts determine the relative phases for the off-axis wakefield structures and, therefore, generate the V-shaped pattern seen in the FREM images. As can be seen in figure \ref{fig:Density} (d), the higher density significantly alters these tilts and thus the phase relationship between the on-axis wakefield and the off-axis wakefields. This is also seen in the FREM image in figure \ref{fig:Density} (h) where the period pattern in the FREM image is altered from those seen in lower densities. The change is also very visible when looking at the longitudinal fields in figure \ref{fig:Density} (i--l). At lower density, as seen is figure \ref{fig:Density} (i), the on-axis field propagates in front of the corresponding off-axis fields. However, this gradually shifts with greater density and by figure \ref{fig:Density} (l), the off-axis fields are in front. 

\subsection{Importance of Ionization on Structure}

Proper consideration of the effect of ionization is critical to the accurate simulation of the flying-focus wakefield. With a standard parabola-generated wakefield, it is often the case that ionization effects, while potentially having a significant impact on electron injection, have relatively minimal impact on the structure of the wakefield itself \cite{Andreev_NonlinearPhenom_2000,Monzac_PRR_2025}. With the axiparabola-generated, flying-focus wakefield, however, the impact is more significant. This is due to the geometry of the rays traveling to the optical axis. The different annular sections of the axiparabola-focused pulse arrive to the axis at different longitudinal positions, so the parts forming the wakefield at the end of the focal line propagate in the gas for a longer distance and are more affected by its presence.
As the field intensity slowly rises as the rays approach the axis, ionization starts to happen at a fairly large radius (hundreds of micrometers) from the axis, and even ionization of the highest levels of atoms (which is often neglected in regular parabola-driven simulations) can affect the propagation of the pulse.

Figure \ref{fig:Ionization} (a--d) show the simulated wakefields generated with axiparabola-focused pulses in the case of fully preionized plasma (a), pure hydrogen (b), pure helium (c), and a 97\% helium--3\% nitrogen mixture for the same conditions as in Figure \ref{fig:experiment}. Figure \ref{fig:Ionization} (e--h) show the simulated FREM images that correspond to the same cases, respectively. Comparing the preionized case (a,e) to the pure hydrogen case (b,f) shows that ionization effects have a minimal impact on the structure of the wake with pure hydrogen. This is reasonable since hydrogen ionizes easily enough that the wakefield forms in plasma that is already fully ionized. 

When comparing the preionized case to that of pure helium (c,g), however, differences can be seen in the wakefield and, especially, in the FREM diagnostic. This means that ionization of the second level of helium which requires higher intensity and happens closer to the axis can have a noticeable additional effect on the pulse propagation. These differences can have a profound impact on the wake's properties, even before any electrons are injected. Therefore, it is critical to properly account for ionization when helium is used. 

Figure \ref{fig:Ionization} (d,h) show the wakefield and FREM simulations for the helium-nitrogen mixture which corresponded to the gas used in the experiment. Interestingly, the additional ionization levels of the nitrogen seems to have a very small impact on the structure of the wake, at least at the 3\% concentration used. Thus, when studying the structure of such flying-focus wakefields, pure helium can be used instead of a mixture. This is significant as PIC simulations of these wakefields already necessitate substantial computational resources and accounting for the ionization of nitrogen can considerably increase the computational load. 

It is important to note, however, that while the effect of the small nitrogen component on the wake structure is limited, the impact on the energy of accelerated electrons can be profound due to different conditions of ionization injection. The influence of nitrogen concentration on the energy of accelerated electrons in flying-focus wakefields was reported in Ref. \cite{Abedi-Varaki_APA_2025}.

\subsection{Importance of Focusing Location}

As was shown in Ref. \cite{liberman2025directobservationwakefieldgenerated}, the structure of the flying-focus generated wakefield evolves significantly over the focal depth. However, the impact of the relative location of the beginning of the focal line to the beginning of the gas target has not previously been investigated. Figure \ref{fig:focusing} (a--d) show simulated wakefields with the same parameters as in figure \ref{fig:experiment} (f) but with different positions $\Delta f_0$ for the start of the focal line relative to the beginning of the density plateau in the simulated gas target. Figure \ref{fig:focusing} (a) has the focal line at the very beginning of the plateau, while figure \ref{fig:focusing} (b), (c), and (d) have the focal line begin 1 mm, 2 mm, and 3 mm past the beginning of the plateau, respectively. Figure \ref{fig:focusing} (e--h) show the FREM images that correspond to these wakefields. 

As can be seen in the FREM images, the change in focusing position has an effect, especially the step between focusing right at the start of the plateau (a) and focusing 1 mm inside (b). This is expected, since for a focus near the beginning of the density profile, almost the entire pulse starts interacting with the gas after crossing the vacuum--gas boundary in the longitudinal direction.
On the contrary, for focusing deeper inside the target, the interaction with the gas begins when the intensity in the transverse direction reaches values sufficient for ionization of the gas, so the value of the relative focusing position becomes less important.

\section{Discussion}

This paper presents a number of insights gained into the behavior of flying-focus wakefields through a combination of direct imaging with femtosecond relativistic electron microscopy and advanced simulations. Experimentally, the stability of the flying-focus wakefield's structure was shown. The simulations, which include optical simulations, particle-in-cell (PIC) simulations, and simulations of the electron probe beam, correspond very closely to the experimental data, strongly implying that they successfully capture the physical dynamics involved. Through these simulations we study the impact of changing the plasma density on the structure of the wakefield and the FREM diagnostic. We show that the wakefield is qualitatively different at higher densities, with the splitting of the laser driver into two distinct pulses, the modification of the off-axis tilts of the peak intensity profile, and the change of the phase between the on-axis and off-axis wakefields. The impact of higher density also presents in a shifting of the structure seen in the FREM images and in a modification of the longitudinal electric field. We also show that, as the density increases, the on-axis longitudinal field distribution becomes more non-linear. Together, these insights shed light onto how flying-focus wakefields could be implemented and imaged at a variety of densities, an important parameter as the ideal gas density for dephasingless acceleration with a flying-focus wakefield spans a large range.

In addition to examining the effect of density, we also study how ionization impacts the structure of the wakefield. We compare the wakefields formed in preionized plasma to the wakefields in initially neutral hydrogen, helium, and helium-nitrogen mixture. We show that ionization has a significant impact on the structure of the wakefield when using a helium target. However, we also show that doping the helium with nitrogen does not significantly alter the wakefield structure. These findings are important for guiding future simulations of the flying-focus wakefield in which the need to accurately capture the physical dynamics coexists with the need to keep the computational demand to feasible level. 

Finally, we examine the influence of the focusing location relative to the gas profile. We show that focusing near the beginning of the gas plateau produces a wakefield that differs significantly from the one obtained when focusing further into the gas jet. This again highlights the impact that the propagation of the axiparabola-reflected laser pulse in the gas has on the properties of the wakefield and the need to optimize the interaction through simulations instead of relying on the properties of the pulse propagation in vacuum.

The insights laid out in this paper will help inform future experiments with flying-focus wakefields. By using the FREM diagnostic to help ensure that our simulations are accurate, we can put increased trust in the understanding that the simulations impart on the behavior of these new wakefields and how it differs from that of the parabola-driven wakefield. Ultimately, this could pave the way towards fully realizing the promise of dephasingless acceleration, and opening up a path towards accelerating electrons with record energies.

\section*{Acknowledgments}
The authors would like to thank Dr. Slava Smartsev for helping develop the experimental setup, Dr. Igor Andriyash for developing the axiprop code, Dr. Eitan Levine for constructive discussions, Prof. Yang Wan for helping develop the code used in the FREM images, and Dr. Eyal Kroupp for helping manufacture the setup. 

\section*{Data Availability Statement}

The raw data used in this work, as well as the the code, can be made available upon reasonable request to the authors. 

\section*{Conflict of Interest}

The authors declare no conflicts of interest.

\section*{Funding}

The research was supported by the Schwartz/Reisman Center for Intense Laser Physics, the Benoziyo Endowment Fund for the Advancement of Science, the Israel Science Foundation, Minerva, Wolfson Foundation, the Schilling Foundation, R. Lapon, Dita and Yehuda Bronicki, and the Helmholtz Association.

\section*{References}

\bibliographystyle{unsrt}
\bibliography{Refs}

\end{document}